\documentclass[sts]{imsart}

\usepackage{algorithm,algorithmic,amsfonts,amssymb,amsmath,amscd,amsthm,bm,latexsym}
\usepackage{natbib}
\usepackage{graphicx}
\usepackage{picins}
\usepackage{upgreek}
\usepackage{units}
\usepackage{nicefrac}
\usepackage[utf8x]{inputenc}

\newcommand\MF{{\mathfrak{M}}}

\newcommand\bx{\mathbf{x}}

\renewcommand{\rho}{\varrho}

\begin{document}

\begin{frontmatter}

\title{The expected demise of the Bayes factor}
\runtitle{Discussion of Ly et al.'s paper}
\thankstext{T1}{Christian P. Robert, CEREMADE, Universit{\' e} Paris-Dauphine, 75775 Paris cedex 16, France
{\sf xian@ceremade.dauphine.fr}. Research partly supported by the Agence Nationale de la Recherche (ANR,
212, rue de Bercy 75012 Paris) through the 2012--2015 grant ANR-11-BS01-0010 ``Calibration'' and by a Institut 
Universitaire de France senior chair. C.P.~Robert is also affiliated as a part-time professor in the Department of
Statistics of the University of Warwick, Coventry, UK. The discussion section was first posted on xianblog.wordpres.com
on Jan 16, 2015. This discussion is dedicated to the memory of Dennis Lindley, who contributed so much to the advance
and formalisation of Bayesian inference during his career and whose kind comments and suggestions were so helpful in
writing \cite{robert:chopin:rousseau:2009}. Quite obviously, this dedication does not aim at claiming support for the
opinions expressed in the paper.}

\begin{aug}
\author{\snm{Christian P.~Robert}}
\affiliation{Universit{\'e} Paris-Dauphine, CEREMADE, University of Warwick, and CREST, Paris}
\end{aug}

\begin{abstract}
This note is a discussion commenting on the paper by Ly et al. on ``Harold Jeffreys's Default Bayes Factor Hypothesis Tests: Explanation,
Extension, and Application in Psychology" and on the perceived shortcomings of the classical Bayesian approach to testing,
while reporting on an alternative approach advanced by \cite{kamary:mengersen:robert:rousseau:2014} as a solution to
this quintessential inference problem.
\end{abstract}

\end{frontmatter}

\section{Introduction}

\begin{quote}
``Jeffreys’s development of the Bayes factor resembles an experimental design for which one studies where the likelihood
functions overlap, how they differ, and in what way the difference can be apparent from the data." 
\end{quote}

The discussion on Harold Jeffreys's default Bayes factor hypothesis tests written by Alexander Ly, Josine Verhagen, and
Eric-Jan Wagenmakers is both a worthwhile survey and an updating reinterpretation {\em cum} explanation of Harold
Jeffreys‘ views on testing. The historical aspects of the paper offer little grip for critical discussion as they stand
true to the unravelling of the testing perspective in Harold Jeffreys' {\em Theory of Probability} (ToP), as a worthy
complement to our earlier exegesis \citep{robert:chopin:rousseau:2009}. I also agree with the focus chosen therein
on the construction of a default solution for the Bayes factor, as this issue is both central to Jeffreys' thinking and to the 
defence of the ``Bayesian choice" in hypothesis testing. My own discussion is therefore mostly written in the

The plan of the paper is as follows: in Section \ref{sec:Og}, I discuss both the presentation made by the authors and
the argumentation coming from Jeffreys about using Bayes factors. The next section presents further arguments against
the use of the Bayes factor, while Section 4 introduces the alternative of a mixture representation put forward by
\cite{kamary:mengersen:robert:rousseau:2014}. Section 5 is a short conclusion.

\section{On ToP and its commentaries}\label{sec:Og}

Ly et al. (2015) starts with a short historical entry on Jeffreys' work and career, which includes
four of his principles, quoted verbatim from the paper:

\begin{enumerate}
    \item ``scientific progress depends primarily on induction";
    \item ``in order to formalize induction one requires a logic of partial belief" [thus enters the Bayesian paradigm];
    \item ``scientific hypotheses can be assigned prior plausibility in accordance with their complexity" [a.k.a., Occam's
razor];
    \item ``classical ``Fisherian" $p$-values are inadequate for the purpose of hypothesis testing".
\end{enumerate}

While I agree with those principles on a general basis, the third principle remains too vague for my own taste and opens
a Pandora box about the meanings of what is simple and what is penalty. (I have had the same difficulty with the call to
Occam's razor principle in other papers like \citealp{jefferys:berger:1992} and \citealp{consonni:forster:2013}.) It is
all very clear to follow such a rule for a one-parameter distribution like the normal $\mathcal{N}(\theta,1)$
distribution, but much less so with ``a" model involving hundreds of parameters and latent variables. I do not think
Harold Jeffreys envisioned at all a general setting of comparing multiple models, in particular because the
$(\nicefrac{1}{2},\nicefrac{1}{2})$ partition of the probability mass has very little to suggest in terms of extensions,
with too many potential alternatives.

\begin{quote}
``Is it of the slightest use to reject a hypothesis until we have some idea of what to put in its place?" H. Jeffreys,
ToP (p.390)
\end{quote}

I obviously support very much the above quote from Jeffreys' ToP, as indeed rejecting a null hypothesis does not sound as an ideal
ultimate goal for statistical inference, but I am also less than convinced about the argument that testing should be
separated from estimation (p.5), even though I recognise the need for defining a separate prior and parameter. My
central difficulty stands with the issue of picking a prior probability of a model, when prior opinions about different
models are at best qualitative and at worst missing.  For instance, when invoking Occam's razor \citep{rasgha2001a},
there is no constructive way of deriving a prior probability $\mathbb{P}(\mathfrak{M}_0)$ for model $\mathfrak{M}_0$.

\begin{quote}
    ``The priors do not represent substantive knowledge of the parameters within the model" H. Jeffreys, ToP (p.13)
\end{quote}

A very relevant point made by the authors in this discussion of ToP is that Harold Jeffreys only considered embedded or
nested hypotheses, a fact that allows for some common parameters between models and hence some form of reference prior,
as argued in \cite{kamary:mengersen:robert:rousseau:2014}. In Jeffreys' ToP setting, it nonetheless seems mathematically
delicate to precise the notion of ``common" parameters, in particular to call for the same (improper) prior on both
parameter sets, as discussed in \cite{robert:chopin:rousseau:2009}. However, the most sensitive issue is, from my
perspective, the derivation of a reference prior on the parameter of interest, which is both fixed under the null and
perspective, the derivation of a reference prior on the parameter of interest, which is both fixed under the null and
unknown under the alternative in ToP. This state of affairs leads to the unfortunate impossibility of utilising improper priors in most
testing settings. Harold Jeffreys thus tried to calibrate the corresponding proper prior by imposing asymptotic
consistency under the alternative and by introducing the notion of ``exact" indeterminacy under ``completely
uninformative" data. Unfortunately, this is not a well-defined concept. That both predictives take the same values for
such ``completely uninformative" data thus sounds more like a post-hoc justification than a way of truly calibrating the
Bayes factor: that any sample with too small a size is ``completely uninformative" is for instance unconvincing. (Why
shouldn't one pick the simplest model by default?) Further, to impose for the Bayes factor to be one for {\em all}
samples with too small a sample size sounds mathematically impossible to achieve in full generality, although two
specific cases are provided in ToP and reproduced in the current paper.  The reconstruction of Jeffreys' derivation of
his reference prior on pp.10-12 of the authors' discussion is quite illuminating of those difficulties (while also
praiseworthy for its clarity). It also shows that the very notion of ``common" parameter cannot be made into a precise
mathematical concept. For instance, if model $\mathfrak{M}_0$ corresponds to the linear regression with a
single covariate $$y=x_1\beta_1+\sigma\epsilon$$ and model $\mathfrak{M}_1$ to the linear regression with an additional
covariate $$y=x_1\beta_1+x_2\beta_2+\sigma\epsilon,$$ except for using the same symbols, there is plenty of room for
arguing against the fact that $(\beta_1,\sigma)$ is ``common" to both models. We certainly expect $\beta_1$ to shrink
as we introduce a secondary explanatory variable, while the variability of the observable around the regression function
should diminish. A further mathematical difficulty with a nested model is that a prior $\pi(\beta_1,\beta_2,\sigma)$ on the
parameters of the embedding model tells us nothing on the embedded model since $\pi(\beta_1,0,\sigma)$ is not defined in a
unique manner \citep{robert:1993b,marin:robert:2010b}.

\begin{quote}``The objective comparison between $\mathcal{M}_+$ and $\mathcal{M}_1$ is then to keep all aspects the same
$\pi_+(\sigma)=\pi_1(\sigma)$." 
\end{quote}

In the normal example, the authors recall and follow the proposal of Harold Jeffreys to use an improper prior
$\pi(\sigma)\propto1/\sigma$ on the nuisance parameter and argue in his defence the quote above. I find their argument
weak in that, if we use an improper prior for $\pi(\sigma)$, the marginal likelihood on the data as given in
(9)---which should not be indexed by $\delta$ or $\sigma$ since both are integrated out---is no longer a probability
density and I do not follow the argument that one should use the same measure with the same constant both on $\sigma$
alone---for the nested hypothesis---and on the $\sigma$ part of $(\mu,\sigma)$---for the nesting hypothesis. Indeed, we
are considering two separate parameter spaces with different dimensions and hence necessarily unrelated measures. Once
again, this quote thus sounds more like wishful thinking than like a genuine justification. 
Similarly, assumptions of independence between $\delta=\mu/\sigma$ and $\sigma$ are not relevant for
$\sigma$-finite measures (even though \cite{hartigan:1983} would object to this statement).  Note that the authors later
point out that the posterior on $\sigma$ varies between models despite using the same data (which somewhat argues
that the parameter $\sigma$ is far from common to both models!). From Jeffreys's perspective, the [testing]
Cauchy prior on $\delta$ is only useful for the testing part and would thus have to be
replaced with another [estimation] prior once the model has been selected [by looking at the data]. This
may thus end up as a back-firing argument about the (far from unique) default
choice. Incidentally, I fail to understand in ToP the relevance of separating (10)
and $s^2 = 0$ from the general case as this former event happens with probability zero,
making Jeffreys' argument at best an approximation to the limiting case of (11).

\begin{quote}
``Using Bayes’ theorem, these priors can then be updated to posteriors conditioned on the data that were actually observed." 
\end{quote}

The re-derivation of Jeffreys' conclusion that a Cauchy prior should be used on $\delta=\mu/\sigma$ highlights the issue
that this choice only proceeds from an imperative of fat tails in the prior, without in the least solving the
calibration of the Cauchy scale, which has no particular reason to be set to $1$. The choice thus involves arbitrariness
to a rather high degree. (Given the now-available modern computing tools, it would be nice to see the impact of this
scale $\gamma$ on the numerical value of the Bayes factor.) And the above choice may also proceed from a ``hidden
agenda", namely to achieve a Bayes factor that solely depends on the $t$ statistic. But this does not sound like a such
compelling reason, given that the $t$ statistic is not sufficient in
this setting.

In a separately interesting way, the authors mention the Savage-Dickey ratio (p.17) as a computational technique to represent the Bayes
factor for nested models, without necessarily perceiving the mathematical difficulty with this ratio that
\cite{marin:robert:2010} exposed a few years ago. For instance, in the psychology example processed in the paper, the test is
between $\delta=0$ and $\delta\ge0$; however, if I set $\pi(\delta=0)=0$ under the alternative prior, which should not
matter [from a measure-theoretic perspective where the density is uniquely defined almost everywhere], the Savage-Dickey
representation of the Bayes factor returns zero, instead of 9.18! The potential for trouble is even clearer in the
one-sided case illustrated on Figure 2, since the prior density is uniformly zero before $\delta=0$ and can be
anything, including zero at $\delta=0$.

\begin{quote}
    ``In general, the fact that different priors result in different Bayes factors should not come as a surprise."
\end{quote}

The second example detailed in the paper is the test for a zero Gaussian correlation. This is a sort of ``ideal case" in
that the parameter of interest is between -1 and 1, hence makes the choice of a uniform U(-1,1) easy or easier to argue.
Furthermore, the setting is also ``ideal" in that the Bayes factor simplifies down to a marginal over the sample
correlation $\hat \rho$ by itself, under the usual Jeffreys priors on means and variances. So we have here a second case where the frequentist
statistic behind the frequentist test[ing procedure] is also the single (and insufficient) part of the data used in the
Bayesian test[ing procedure]. Once again, we thus are in a setting where Bayesian and frequentist answers are in one-to-one
correspondence (at least for a fixed sample size) and where the Bayes factor allows for a closed form through
hypergeometric functions, even in the one-sided case. (This is a result obtained by the authors, not by Harold Jeffreys who, as
the proper physicist he was, obtained approximations that are remarkably accurate!)

\begin{quote}
    ``The Bayes factor (...) balances the tension between parsimony and goodness of fit, (...) against overfitting the data."
\end{quote}

{\em In fine}, I liked very much this re-reading of Jeffreys' approach to Bayesian testing, maybe the more because I now
consider we should move away from this approach as discussed below. However, I am not certain the discussion will help
in convincing psychologists to adopt Bayes factors for assessing their experiments as it may instead frighten them away.
And as it does not bring an answer to the vexing issue of the relevance of point null hypotheses. But the paper
constitutes a lucid and innovative treatment of the major advance represented by Jeffreys' formalisation of Bayesian
testing.

\section{Further misgiving about the Bayes factor}

In this section, I extrapolate on some difficulties I have with the Bayes factor, as discussed in more depth in
\cite{kamary:mengersen:robert:rousseau:2014}.

The natural Bayesian decision-theoretic approach to decide between two models is to use a binary $0-1$ loss function and
to derive the posterior probabilities of the model. However, since this approach depends on the choice of 
unnatural prior weights, Harold Jeffreys advocates in ToP its replacement with Bayes factors that eliminate this dependence.
Unfortunately, while indicative of the respective supports brought by the data through their comparison with 1, Bayes
factors escape a direct connection with the posterior distribution, for the very reason they eliminate the prior weights. Therefore,
they lack the direct scaling associated with a posterior probability and a loss function. This implies that
they face a subsequent and delicate interpretation (or calibration) and explains why ToP does not contain a section on
decision making using Bayes factors, instead providing in the Appendix a logarithmic scale of strength that is purely
qualitative. Note that posterior probabilities face similar difficulties, in that they suffer from the unavoidable tendency to interpret 
them as $p$-values and to scale them against the 5\% ~reference value when {\em de facto} they only report through a
marginal likelihood ratio the 
respective strengths of fitting the data to both models. (This difficulty is not to be confused with the divergence in
the [frequentist versus epistemic] interpretations of the probability statement, as discussed in \cite{fraser:2011} and
the ensuing comments.)

At a different semantic level, the long-going [or long-winded] criticism on the Bayesian approach, namely the dependence
on a subjective prior measure applies here twofold: first in the linear impact of the prior weights of the models under
comparison and second in the lasting impact of the prior modelling on the parameter spaces of both models under
comparison. (We stress as in \cite{robert:2014} a rather overlooked feature answering such criticisms (see, e.g.,
\citealp{spanos:2013}), including the
Lindley-Jeffreys (\citeyear{lindley:1957}) paradox, namely the overall consistency of Bayes factors.) However, the
resulting discontinuity in the use of improper priors is a feature I have always been uncomfortable with, since those
improper priors are not justified \citep{degroot:1982} in most testing situations,
leading to many alternative if {\it ad hoc} solutions \citep{robert:2004}, where data is either used twice
\citep{aitkin:2010} or split in artificial ways \citep{berger:pericchi:1996,berger:perrichi:1998}.

As pointed out in the above, the posterior probability is more often associated with a binary ({\it accept}
vs.~{\it reject}) outcome that is more suited for immediate decision (if any) than for model evaluation, in connection with the
rudimentary loss function behind it, while the Bayes factor is a smoother form of comparison that should not {\em in
fine} conclude with a (hard) decision. Given the current abuse of $p$-values and significance tests
\citep{johnson:2013b}, we should advocate more strongly this stand. In conjunction with this point, note further that
returning a posterior probability or a Bayes factor offers no window into the uncertainty associated with the decision
itself, unless one proceeds through additional and most likely costly simulation experiments.

From a computational perspective, let me recall there is no universally acknowledged approach to compute marginal
likelihoods and Bayes factors \citep{chen:shao:ibrahim:2000,marin:robert:2010}, while some approaches are notoriously
misleading \citep{newton:raftery:1994}. In a recent work about the validation of ABC model selection
\citep{robert:cornuet:marin:pillai:2011,marin:pillai:robert:rousseau:2011}, we also pointed out the variability of the
numerical estimates and {\em in fine} the utter dependence of both posterior probabilities and Bayes factors on conditioning 
statistics, which in turn undermines their validity for model assessment.

\section{Testing as (mixture) estimation}

The alternative to testing via Bayes factors, as proposed in \cite{kamary:mengersen:robert:rousseau:2014}, to which the
reader is referred to for details, constitutes
a paradigm shift in the Bayesian processing of hypothesis testing and of model selection in that it reformulates both
the question and the answer into a new framework that accounts for uncertainty and returns a posterior distribution
instead of a single number or a decision. As demonstrated in \cite{kamary:mengersen:robert:rousseau:2014}, this shift
offers convergent and naturally interpretable solution, while encompassing a more extended use of improper priors.
The central idea to the approach is to adopt a simple representation of the testing problem as a two-component mixture
estimation problem where the weights are formally equal to $0$ or $1$ and to estimate those weights as in a regular
mixture estimation framework. This approach is inspired from the consistency results of \cite{rousseau:mengersen:2011} 
on estimated overfitting mixtures, i.e., mixture models where the data is actually issued from a mixture distribution
with a smaller number of components.

More formally, given two statistical models,
$$
\MF_0:\ x\sim f_0(x|\theta_0)\,,\ \theta_0\in\Theta_0 \quad\text{and}\quad
\MF_1:\ x\sim f_1(x|\theta_1)\,,\ \theta_1\in\Theta_0\,,
$$
\cite{kamary:mengersen:robert:rousseau:2014} define the (arithmetic) encompassing mixture model
\begin{equation}\label{eq:mix}
\MF_\alpha:\ x\sim \alpha f_0(x|\theta_0) + (1-\alpha) f_1(x|\theta_1)\,,
\end{equation}
with a mixture weight $0\le \alpha\le 1$, meaning that each element of the iid sample associated with the model comparison is considered
as generated according to $\MF_\alpha$. While this new and artificial model contains or encompasses both $\MF_0$ and $\MF_1$ as two special
cases, that is, when $\alpha=0$ and $\alpha=1$, a standard Bayesian analysis of the above mixture provides an estimate
of the weight $\alpha$, relying on a prior distribution $\pi(\alpha)$ with support the entire $(0,1)$ interval, e.g., a
Beta $\mathcal{B}e(a_0,a_0)$ distribution. This means that such a standard processing of the model will create a
posterior distribution on the weight $\alpha$,  given the data, which location on the unit interval will induce evidence
(and strength of evidence) in favour of one model versus the other. For instance, when this posterior is concentrated
near zero, the data supports more strongly $\MF_1$ than $\MF_0$. Hence, this alternative paradigm does not return a
value in the binary set $\{0,1\}$ as a more traditional decision strategy or a test would do.  Thus, the mixture
representation is quite distinct from making a choice between both models (or hypotheses) or even from computing a
posterior probability of $\MF_0$ or $\MF_1$. Inference on the mixture representation bypasses the testing difficulties
produced in the previous section in that there is no decision involved.  I thus consider it has the potential of a more
natural and easier to implement approach to testing, while not expanding on the total number of parameters when compared
with the original approach (as found in ToP). \cite{kamary:mengersen:robert:rousseau:2014} argue in favour of this
shift from several perspectives, ranging from inferential to computational, and I once again refer to this paper for further
implementation details, consistency proof, and examples.  

While the encompassing mixture model $\MF_\alpha$ intrinsically differs from the true model, given that the weight
$\alpha$ can take any value in $(0,1)$, the production of a posterior distribution on the weight $\alpha$ must be deemed
to be a positive feature of this completely different approach to testing in that it bypasses the vexing issue of setting 
artificial prior probabilities on the different model indices and that it measures
the proximity of the data to both models by a probability to allocate {\em each} datapoint to those models. Furthermore,
the mixture model $\MF_\alpha$ allows for the use of partially improper priors since both components may then enjoy
common parameters, as for instance location and scale parameters. This feature implies that using the {\em same}
reference measure on the nuisance parameters of both models is then absolutely valid. Estimating a mixture model by MCMC
tools is well-established \citep{diebolt:robert:1990a,celeux:hurn:robert:2000} and bypasses the difficulty in computing
the Bayes factor.  At last, an approach by mixture modelling is quite easily calibrated by solutions like the parametric
bootstrap, which that there is no decision involved. I thus consider this novel formalism has the potential of a better
approach to testing, while not expanding on the number of parameters when compared with the original approach (as found in ToP).
\cite{kamary:mengersen:robert:rousseau:2014} argue in favour of this shift from several perspectives, from inferential
to computational, and I once again refer to this paper for further details.

While the encompassing model $\MF_\alpha$ intrinsically differs from the real model, given that the weight $\alpha$ can take
any value in $(0,1)$, the production of a posterior distribution on the weight $\alpha$ is a positive feature of this
approach in that it bypasses the vexing issue of setting artificial prior probabilities on model indices and measures
the proximity of the data to the models. Furthermore, $\MF_\alpha$ allows for the use of partially improper priors since
both components may then enjoy common parameters, as for instance location and scale parameters. This implies that using
the same reference measure on the nuisance parameters of both models is then completely valid.
At last, the approach by mixture modelling is quite easily calibrated by solutions like the parametric bootstrap, which
provides a reference posterior of $\alpha$ under each of the models under comparison.

\begin{figure}[!h]
\includegraphics[width=.59\textwidth]{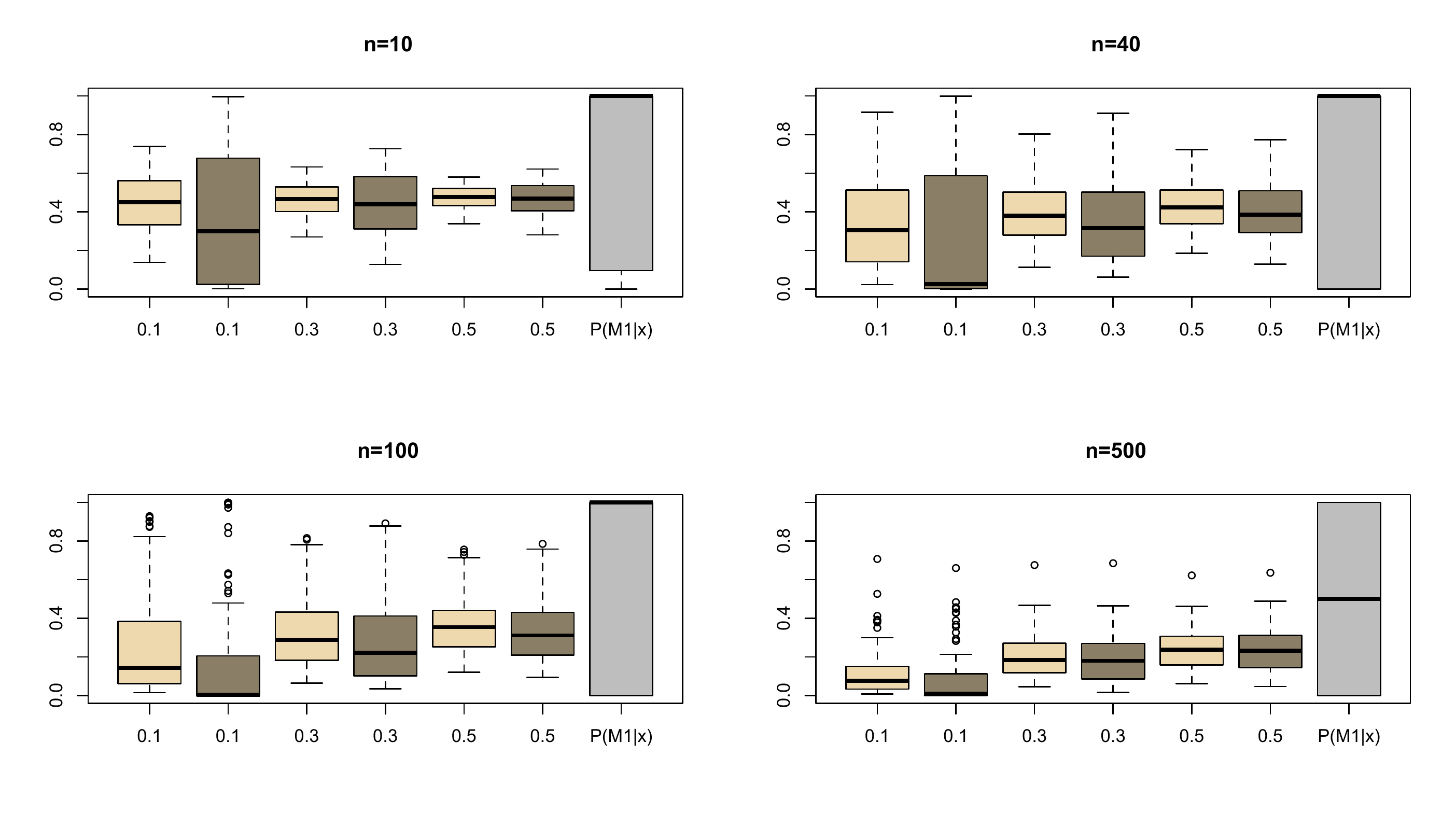}\includegraphics[width=.41\textwidth]{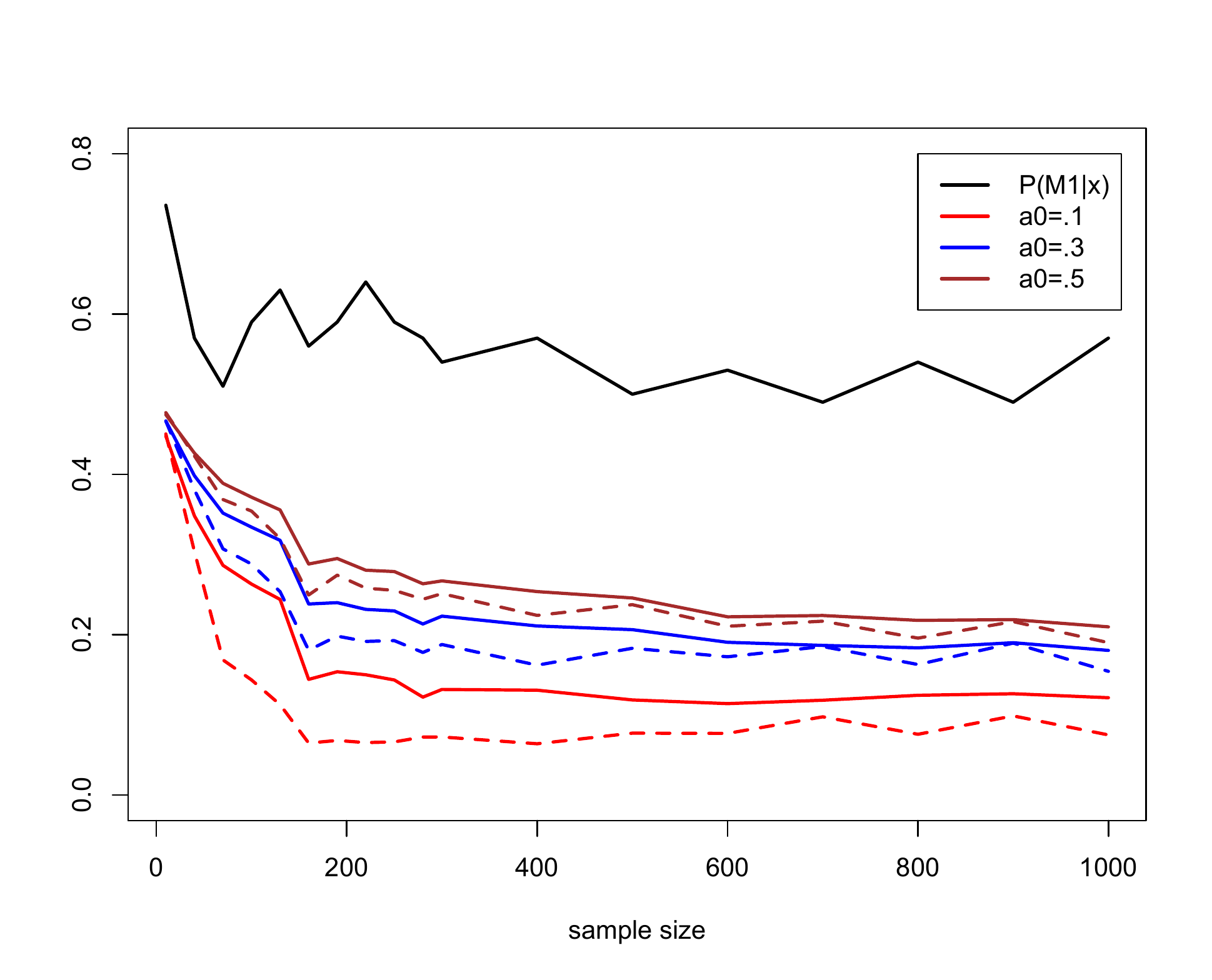}
\caption{\small {\em (left)} Boxplot of the posterior means {\em(wheat)} and medians {\em(dark wheat)} of
the mixture weight $\alpha$, and of the posterior probabilities of model $\mathcal{N}(\mu, 1)$ {\em(blue)} evaluated over 100 replicas of
$\mathcal{N}(0, .7^2)$ datasets with sample sizes $n=10, 40, 100, 500$; {\em (right)} evolution across sample sizes
of the averages of the posterior means and posterior medians of $\alpha$, and of the posterior probabilities $\mathbb{P}(\MF_0|\bx)$, 
where $\MF_0$ stands for the $\mathcal{N}(\mu,1)$ model. Each posterior estimation of
$\alpha$ is based on $10^4$ Metropolis-Hastings iterations. {\em [Source:
\citealp{kamary:mengersen:robert:rousseau:2014}, with permission.]}}
\label{fig:nono}
\end{figure}

From a practical perspective---even though it involves a paradigm shift from the current practice of referring to a gold
standard, like $0.05$---, the implementation of the principle of \cite{kamary:mengersen:robert:rousseau:2014} means
estimating the mixture model by a computational tool like MCMC and exploiting the resulting posterior distribution on
$\alpha$ in the same way any posterior is to be interpreted. Rather than advocating hard decision bounds associated with
such a posterior $\pi(\alpha|\mathcal{D})$, as in an alternative $p$-value with similar drawbacks
\citep{ziliak:mccloskey:2008}, it is more natural to contemplate the concentration of this distribution
near the boundaries, $0$ and $1$, in absolute terms and relative to the concentration of a posterior associated with a
sample from either model. For a sample size that is large enough, this concentration should be clear enough to conclude
in favour of one model. Figure \ref{fig:nono} illustrates this approach, for normal samples of sizes ranging from $n=10$
to $n=500$, when opposing the point null $\mathcal{N}(0,1)$ model ($\MF_1$) to the alternative $\mathcal{N}(\mu,1)$ model
($\MF_0$), under a proper $\mu\sim\mathcal{N}(0,1)$ prior. As can be inferred from the left panel, the posterior estimates of $\alpha$, whether
posterior means or posterior medians, concentrate faster with $n$ on the relevant boundary, that is, close to zero and
in favour of $\MF_0$, than the exact posterior probability (associated with a prior weight of $\nicefrac{1}{2}$ on both
models, and hence allows us to conclude more quickly about the evidence in favour of the null model. As seen from the
right panel, the impact of the hyper-parameter value in the prior modelling $\alpha\sim\mathcal{B}e(a_0,a_0)$ remains moderate.

\section{Conclusion}

\begin{quote}
    ``In induction there is no harm in being occasionally wrong; it is inevitable that we shall be." H. Jeffreys, ToP (p.302)
\end{quote}

As a genuine pioneer in the field, Harold Jeffreys\footnote{Whose above quote may be a pastiche of Keynes' own 1933 ``There
is no harm in being sometimes wrong -- especially if one is promptly found out".} set a well-defined track, namely the Bayes factor, for conducting Bayesian
testing and by extension model selection, a track that has become the norm in Bayesian analysis, while incorporating the
fundamental aspect of reference priors and highly limited prior information. However, I see this solution as a child of
its time, namely, as impacted by the on-going formalisation of testing by other pioneers like Jerzy Neyman or Egon
Pearson. Returning a single quantity for the comparison of two models fits naturally in decision making, but I strongly
feel in favour of the alternative route that Bayesian model comparison should abstain from automated and hard decision making.
Looking at the marginal likelihood of a model as evidence makes it harder to refrain from setting decision bounds when
compared with returning a posterior
distribution on $\alpha$ or an associated predictive quantity, as further discussed in
\cite{kamary:mengersen:robert:rousseau:2014}. 

Different perspectives on this issue of constructing reference testing solutions are obviously welcome, from the
incorporation of testing into the PC priors and baseline models of \citep{simpson:etal:2014} to the non-local tests of
\cite{johnson:rossell:2010}, and I would most gladly welcome exchanges on such perspectives.

\section*{Acknowledgements}
I am quite grateful to Kerrie Mengersen (QUT) and Joris Mulder (Tilburg University) 
for helpful comments and suggestions on this paper. Discussions with the authors of the
paper during a visit to Amsterdam and their kind welcome are also warmly acknowledged.

\hyphenation{Post-Script Sprin-ger}

\end{document}